\documentstyle[12pt]{article}
\begin{document}
\Large
\begin{center}
{\bf                   Gluon exchange and $p_{T}$ distribution} \\
\vskip 0.5cm
\large
                       Vladimir Lugovoi
\vskip 0.1cm
{\small  \em           Department of High Energy Physics, Physical Technical Institute
                       of Tashkent , Mavlyanov st. 2b , Tashkent , \rm 700084 \small \em Uzbekistan}
\end{center}
\vskip 0.1cm
\begin{center}
{ \normalsize            Electronic addresses : lugovoi@lug.post.tps.uz

 $\;\;\;\;\;\;\;\;\;\;\;\;\;\;\;\;\;\;\;\;\;\;\;\;\;\;\;\;\;\;\;\;\;\;\; $
                                                lugovoi@physic.uzsci.net }
\end{center}
\vskip 0.7cm
\normalsize

\begin{abstract}

 It is shown that even if a quark $q$ is on the mass shell, the quark $q'$
created via vacuum pair production in the process $q \rightarrow h + q'$
has wide distribution for its spacelike virtuality, and so at the end of
daughter string $(q'\overline{q})$ there is a parton having wide distributed
spacelike virtuality.  However, the essence of the parton model is to regard
a high energy hadron-like cluster as a collection of quasifree partons.
Therefore we use a set of some principles of string model, formulated,
for example, by Andersson and Nilsson, that gluons move like localised
particles carrying energy momentum between quark and antiquark at the
endpoints of string, and so we take into account they go on the mass shell
before the break of daughter string occurs.  It is stressed, the gluons
carry also transverse momentum which could be connected with spacelike
virtuality of quark at the end of string. Some approximations to describe the
gluon exchange are studied and results of Monte Carlo calculations for
antiproton-proton interactions are given.
\end{abstract}
\newpage
\large

\section{Introduction}

Well known concepts of a quark confinement and quantum mechanical
tunneling of quark-antiquark pair from the vacuum are used actively in
many models of multiparticle production \cite{Artru}-\cite{Amelin}.
An example of quantitative description of quark confinement was given in the models
\cite{Artru}-\cite{Werner} where the evolution in space and
time for quark and antiquark at the endpoints of string occurs according to
classical string theory and the quark-antiquark colourless system is confined
by constant magnitude force which determines the breakpoint of string.
The other description for evolution of same colourless system of confined quarks
is given in the models \cite{AnderGustIngSjos}-\cite{Amelin} where already
the momentum energy representation and scaling distributions\footnote{i.e. these
distributions do not depend on the energy of the (anti)(di)quark at the
endpoint of string.} are used to describe the string breaking.
In this fragmentation procedure there is the parton approach,
but in that way a quark confinement means also that an interaction
has to be between the quark and antiquark at the endpoints of string.
So, we are going now to construct a Monte Carlo model which takes into account
a quantitative description for interaction between the quark and antiquark
at the  endpoints of string according to concepts of the models
 \cite{AnderGustIngSjos}-\cite{Amelin}.
It has to be stressed, our new point of view on the process is that
we suppose the interaction between the both
endpoints of string precedes each break of string into daughter string and
hadron\footnote{or into two daughter hadrons}.

Let us remind the Monte Carlo string breaking scheme exploited in well-known
models \cite{AnderGustIngSjos}-\cite{Amelin} where a conversion of quark $q$,
which is endpoint of primary string, into hadrons $h$ (Fig.1a) is an iteration
of an elementary process presented in Fig.1b : quark $q$ is converted into
hadron $h$ and quark $q'$ via the vacuum pair production. Therefore the produced
$q'$ does not lie on the mass shell\footnote{In the Monte Carlo models
\cite{AnderGustIngSjos}-\cite{Amelin} the energy and momentum vector of
produced quark $q'$ are determined as the differences between energies and
momentum vectors of $q$ and $h$.}, and so in the next conversion
presented in Fig.1c ($q' \rightarrow h + q''$) the virtuality\footnote{In
the models \cite{AnderGustIngSjos}-\cite{Amelin} there is not conversion
of quark (into hadron and quark), if it has the negative energy.
Therefore we can choose the quark mass squared invariant
$m_{q}^{* 2} = E^{2}_{quark}- {\vec{P}}^{2}_{quark}$ as a quantitative
characteristic for virtuality of quark.} of quark
$q''$ rises above the virtuality of quark $q'$, i.e. the virtuality
of each produced quark grows with the number of
conversions presented in Fig.1a.   It does  not take into account in the models
\cite{AnderGustIngSjos}-\cite{Amelin} where the scaling distributions for the
fraction $z$ of the longitudinal and transverse momentum projections carryed
by hadron\footnote{In the model \cite{AurenchBoppRanft} the fraction of
longitudinal end transverse energy carryed by hadron $h$ are generated
instead of fraction of longitudinal end transverse momentum.} $h$ do not
depend on  virtuality of quark $q$ (Fig.1b) converted into daughter hadron $h$
and quark $q'$. Hovewer, as it is shown in Fig.2, the produced quark $q'$
(diquark $qq'$)  has wide distribution for its  mass squared invariant
$m_{q'(qq')}^{* 2} = E^{2}_{q'(qq')}- \vec{P}^{2}_{q'(qq')}$
in negative region even if quark $q$ (see Fig.1b) lies on the mass shell
in each its conversion into daughter hadron $h$ and quark $q'$ (diquark $qq'$).
For example, if quark $q$ is on the mass shell in each conversion
(Fig.1b), the average values of the quark mass squared invariants $m_{q'(qq')}^{* 2}$
for daughter $u$-quark, $s$-quark and diquark produced in the processes
$q\rightarrow h+q'_{u}$ , $q\rightarrow h+q'_{s}$ and $q\rightarrow h+qq'$
are  $(-0.724 \pm 0.009)GeV^{2}$,
$(-1.020 \pm 0.012)GeV^{2}$ and $(-1.940 \pm 0.022)GeV^{2}$
respectively\footnote{These numbers will move farther into negative region,
if quark $q$ does not lie on the mass shell in the conversion
$q \rightarrow h + q'$.}.

Furthermore, in the fragmentation schemes of the Monte Carlo models
\cite{AnderGustIngSjos}-\cite{Amelin} two partons at the ends of primary
string are converted into two jets of hadrons independently (Fig.1d),
i.e. any interaction between partons at the ends of each daughter string
does not take into account.

Therefore  we try to describe here a Monte Carlo
version for these interactions included in the model without contradiction
to a set  of concepts of the string model formulated by B.Andersson and
A.Nilsson in the paper \cite{AnderNils}, which are :
\vskip 0.4cm
{\em quarks, $q$, and antiquarks, $\overline{q}$, are treated as excitations at the

endpoints of the force field and gluons, $g^{,}$s, are internal exci-

tations. The massless relativistic string is used as a (semi-

classical) model for the force field. The above-mentioned exci-

tations move like localised particles carrying energy momen-

tum.}
\vskip 0.4cm
Even for that quasi-classical approximation the string interaction is realised
by gluons, $g^{,}$s, which move like localised particles carrying energy
momentum between the endpoints of string. It is like a classical
parton model approximation\footnote{where quarks and gluons are localised
particles carrying some parts of energy-momentum of hadron} and it gives us
a way to construct a some model for confinement  interaction.
Therefore let us suppose the following scenario for confinement interaction
and string breaking. The endpoints of string are
quark, $q'$, and antiquark, $\overline{q}$. They interact each other via gluon
exchange. The gluons, $g^{,}$s, carry energy momentum, so the energy momentum
of partons at the endpoints of each daughter string is changed. As a result
both partons turn out to be on the mass shell\footnote{If one of them had been
produced from the vacuum ($q'$ in Fig.1b), it would not have lain on
the mass shell (Fig.2) before gluon excange (circle in Fig.3b) occurs between
both partons ($q'$ and $\overline{q}$ in Fig.3b) at the endpoints of string.}
that  allows to find their new energies and momenta in the rest frame of the
string where then a conversion (for example\footnote{The conversions of quark
$q' \rightarrow h + q''$ and antiquark
$\overline{q} \rightarrow h + \overline{q}'$ at the endpoints of string
are generated with equal probabilities.}) of quark $q' \rightarrow h + q''$
is generated. This algorithm for produced daughter string tensed already
by antiquark $\overline{q}$ and quark $q''$ produced from vacuum is
repeated up to the mass of string becomes low (see Sect.4) and decay of string
into two hadrons occurs.  In frame of this scenario (anti)(di)quark
(parton at the endpoint of string) turns out
to be on the mass shell before it is converted into hadron and quark (Fig.1b).
So, there is not any problem connected with virtuality
and simultaneously we take into account the interaction
between quark and antiquark at the endpoints of each daughter string
produced during cascade decay (Fig.3b) of primary string.

The mechanism for string breaking is visioned via the tunneling
of $q'$-$\overline{q}'$ pair from vacuum (Fig.1b). Usually the Gauss law is
used to describe the distribution for the transverse momentum projections
$p_{x}$ ,  $p_{y}$ of vacuum
quark. So, if we use the Gaussian distribution  which has zero center
parameter\footnote{i.e. the summary transverse
momentum of the vacuum quark-antiquark pair is equal to zero.}, we have to
generate $p_{x}$ and $p_{y}$ of vacuum quark $q'$ in the system
where the transverse momentum of converted quark $q$ (Fig.1b) is
zero\footnote{it is the $K_{decay}$ frame determined in Sect.4}.
To generate the $x$- and $y$-projections for $u(d)$-quark, $s$-quark and diquark
from vacuum pair we use, for Gaussian distribution, the widths
\begin{equation}
\sigma_{u(d)}^{vac}    = 0.15 \; GeV ,   \;\;\;\;
\sigma_{s}^{vac}       = 0.25 \; GeV ,   \;\;\;\;
\sigma_{diquark}^{vac} = 0.35 \; GeV
\end{equation}
respectively.

The other process is gluon exchange between the quarks
at the endpoints of string. The gluons carry also
the transverse momentum $p_{T}$.
The Gauss law usually is used to describe the distributions of the $p_{T x}$
and $p_{T y}$ projections in the non-perturbative process,
if the transverse momentum exchanges are considered as uncorrelated.
However the string interaction is gluon exchange between partons having
long tail distributions for (di)quark mass squared invariant
$m_{q'(qq')}^{* 2}$  (see Fig.2).
Therefore if we use the Gauss law to descbibe $p_{T}$ distribution for the
gluon exchange process\footnote{in string rest frame
where quark and antiquark at the endpoints of string have zero
transverse momenta and move in opposite directions (it is $K_{exchange}$
frame determined in Sect.3).},
the Gauss width parameter might be connected with spacelike virtuality
of partons at the endpoints of string.    It seems to be reasonable.
Namely, gluon exchange between partons, having big
spacelike virtuality (Fig.2), can lead to different $p_{T}$ because after
interaction each other the both partons turn out to be on the mass shell.
However, the essence of the parton model is to regard
a high energy claster (hadron, string) as a collection of quasifree partons.
So, if both partons at the ends of string are on the mass shell,
the transfered $p_{T}$ of gluons has to be zero.  The variable which
has the similar properties is the
difference between the quark mass squared invariant
$m_{q}^{* 2} = E^{2}_{quark}- {\vec{P}}^{2}_{quark}$ and
the fixed quark mass squared
$m_{q}^{2}$ taken from  \cite{AnderGustIngSjos}
at the given type of quark $q$. This difference equals to zero
(i.e. $m_{q}^{* 2} = m_{q}^{2}$), if quark is on the mass shell,
and so the variable\footnote{Here $q$ is parton which does not lie
on the mass shell before gluon exchange occurs.}
\begin{equation}
 \sigma_{q}^{} \;\; = \;\; \mid m_{q}^{* 2}-m_{q}^{2}\mid^{0.5}
\end{equation}
could be chosen as the width parameter for Gaussian distribution of
$x$-, $y$-projections of parton transverse momentum after gluon exchange.
Of course, this approximation is discussionable, so in our calculations
we chose more simple approximation. Instead of the variable
$m_{q}^{* 2}$  we took its average values $<m_{q}^{* 2}>$ given above.
Namely, instead of eq.(2) we used
\begin{equation}
\sigma_{q} \;\; = \;\; \mid <m_{q}^{* 2}>-m_{q}^{2}\mid^{0.5} \;\;\;,
\end{equation}
where $<m_{q}^{* 2}>$ equals to $(-0.724 \pm 0.009)GeV^{2}$,
$(-1.020 \pm 0.012)GeV^{2}$ and $(-1.940 \pm 0.022)GeV^{2}$
for $u(d)$-quark, $s$-quark and diquark respectively. Now from eq.(3) one can
get\footnote{A (di)quark at the endpoint of strig occurs from the vacuum pair
where usual relation between quark flavours is $u:d:s = 3:3:1$,
and so the average mass of diquark is 0.49 GeV. To calculate this
average value we put $m_{ud}=0.45 GeV$,
$m_{ds}=0.62 GeV$, $m_{ss}=0.79 GeV$ \cite{AnderGustIngSjos}.}
\begin{equation}
    \sigma_{u(d)}    = 0.92 \; GeV,    \;\;\;\;
    \sigma_{s}       = 1.13 \; GeV,    \;\;\;\;
    \sigma_{diquark} = 1.48 \; GeV     \; .
\end{equation}
These values, describing the $p_{T}$ distribution of quark after gluon exchange,
are in a few times more than the same values (see (1)) describing
the $p_{T}$ distribution of quark from vacuum pair.
Therefore in our calculations we put the relations
\begin{equation}
 \sigma_{u(d)}       =  3 \cdot \sigma_{u(d)}^{vac}     \;\;\;\;, \;\;\;
 \sigma_{s}          =  3 \cdot \sigma_{s}^{vac}        \;\;\;\;, \;\;\;
 \sigma_{diquark}    =  3 \cdot \sigma_{diquark}^{vac}  \;\;.
\end{equation}

\section{         The formation of primary strings     }

The initial stage of soft\footnote{In the presented model we do not take into
account the (semi)hard interactions between the partons of projectile and
target.} hadron-hadron collision is the formation of colourless strings.
In our calculations we use the Dual Parton Model (DPM)
\cite{Kai-Ter-Mar}-\cite{Capella} to forme the primary strings.  In DPM
the primary colourless strings are stretched by partons of target
and projectile (see Fig.4).   In our model the number $l$ of pair of primary
strings is generated firstly according to theoretical distribution
 \cite{Kai-Ter-Mar} (see Appendix 1).
The flavours of sea quarks in projectile and target (see Fig.4) and
the flavours of quarks, produced from vacuum at the decay of string (see Sect.4),
are generated according to relation
\begin{equation}
u : d : s = 3 : 3 : 1  .
\end{equation}
The initial multidimensional distribution of partons on the part $x_{i}$
of longitudinal momentum of projectile in the centre mass system is given by
\cite{Cap}, \cite{Capella}

\begin{equation}
x_{1}^{-0.5} \; ( x^{2}_{2}    + \mu^{2}_{2}    / P^{2} )^{-0.5} \; ...
             \; ( x^{2}_{2l-1} + \mu^{2}_{2l-1} / P^{2} )^{-0.5} \;
                  x^{1.5}_{2l}
\;\; dx_{1} \; dx_{2} \; ... \; dx_{2l}
\end{equation}
at $0\leq x_{i} \leq 1$ , where $P$ is the momentum of proton in c.m.s.;
indeces $1$ and $2l$ are related to the valence quark and diquark correspondingly,
the remaining indices are related to sea quarks and antiquarks;
$\mu_{i}$ is the mass of $i$-th quark or antiquark ($\mu_{u}$ = $\mu_{d}$ = 0.34 GeV,
$\mu_{s}$ = 0.51 GeV). The $x'_{i}$-values of target partons are generated
identically. For realization of longitudinal momentum conservation after
$x_{i}$ and $x'_{i}$ generation we substitute

\begin{equation}
 x_{i} / \sum_{j=1}^{2l} x_{j} \;\; \rightarrow \;\;   x_{i} \;\;\; \;\;\;,\;\;\;\;\;\;
x'_{i} / \sum_{j=1}^{2l} x'_{j} \;\; \rightarrow \;\;   x'_{i} \;\;\; .
\end{equation}
Taking into consideration the energies of projectile quark (antiquark)
$\varepsilon_{i} = ( \mu^{2}_{i} + x^{2}_{i} P^{2} )^{0.5}$
and target quark (antiquark)
$\varepsilon'_{i} = ( {\mu'_{i}}^{2} + {x'_{i}}^{2} P^{2} )^{0.5}$ ,
one can find the energy $E_{i}$ , longitudinal momentum $P_{i}$ and
mass  $M_{i}$ of every $i$-th string. The transverse momenta of partons are not
accounted for. Then we control for every $i$-th string the fulfilment of
inequalities

\begin{equation}
E_{i} > |P_{i}| \;, \;\;\;  M_{i}  >  \mu_{i1} \;\; +
\;\; \mu_{i2} \;  , \;\;\;  M_{i}  >  M_{i1}^{min} \;\; +
\;\; M_{i2}^{min} \; ,
\end{equation}
where $ \mu_{i1} $ and $ \mu_{i2} $ are masses of the partons,
stretching the string, and the $M_{ij}^{min}$ is minimum mass of
the $j$-th hadron  with the fixed quark composition,
which can be produced by decay of the $i$-th string\footnote{See Sect.4.}.
The Monte Carlo event is rejected, if only one inequality in eq.(9) is not
fulfiled. The share  of rejected events depends on the $l$ value,
so the distribution of unrejected events over $l$ differs from
the initial distribution \cite{Kai-Ter-Mar}.
Hovewer the share of rejected MC events grows with
the number $l$ of primary strings. Therefore the difference between the
theoretical distribution and the distribution for Monte Carlo events on $l$
is only for large $l$ which have a little share (about a few percents)
in theoretical distribution. So, for a first approximation
the both distributions are same\footnote{For example, the share of
rejected Monte Carlo events is 0.10 at $\sqrt{s}$=27 GeV and 0.03 at
$\sqrt{s}$=540 GeV. } each other.
\vskip 1.0 cm

\section{Gluon exchange}

There are many feasible ways to calculate the gluon exchange between both
partons at the ends of string. One of them is given in this Section.

Before every break, for example, of secondary string of mass $M$, occurs,
we know following its characteristics (in the $K_{0}$ c.m.s. of
antiproton-proton collisions): the scheme of the decay of secondary string
(see  Sect.4 and Figs.3,5), the flavours of the partons participating in this decay,
the mass $M$, the energy $E_{0}$ and the momentum vector $\vec{P_{0}}$
of the string and
the energies $\varepsilon_{01}$, $\varepsilon_{02}$ and momentum vectors
$\vec{p}_{01}$, $\vec{p}_{02}$ of first and second partons\footnote{One of them
can have a big specelike virtuality.}, stretching the string\footnote{Of course,
the energies and momentum vectors of partons determine the mass $M$, the energy
$E_{0}$ and the momentum vector $\vec{P}_{0}$ of the string.}.

As it was discussed in Sect.1, it is convenient way to generate the gluon
exchange in the string rest frame. So, the kinematical characteristics of
both partons
should be transformed from $K_{0}$ to string rest frame. It is described
below in details :

\vskip 0.8 cm
{\em Transformation} from $K_{0}$ to $K_{L}$ frame where the longitudinal momentum
$P_{z0}$ of string is equal zero :
\vskip 0.2 cm

This $K_{L}$ frame moves with $\beta_{z0} = P_{z0} / E_{0}$ velocity along the
$z_{0}$ axis in the  $K_{0}$ frame and with Lorentz-factor
$\gamma_{0}=E_{0}/M_{\perp}=E_{0}/(E_{0}^{2}-P_{z0}^{2})^{0.5} $ , where
$M_{\perp }=(M^{2}+P_{\perp 0 }^{2})^{0.5} $ is transverse mass of
string.

\vskip 0.8 cm
{\em Transformation} from $K_{L}$ to $K'$ frame where the longitudinal momentum
$P_{z0}$ of string is equal zero and $x'$ axis is along the transverse
momentum vector of string :
\vskip 0.2 cm

In the $K_{L}$ frame, the string has transverse momentum
$\vec{P}_{\perp 0}$ with projections $P_{x0}$ and $P_{y0}$.
We turn the $K_{L}$ frame (in the transverse $xy$ plane) to obtain
the $x_{L}$ axis along the vector $\vec{P}_{\perp 0}$.
The parameters for this transformation are
$ \cos\varphi_{\perp} = P_{x0}/P_{\perp 0} $,
$ \sin\varphi_{\perp} = P_{y0}/P_{\perp 0} $.

\vskip 0.8 cm
{\em Transformation} from $K'$ to $K_{rest}$ frame where the momentum
of string is equal zero :
\vskip 0.2 cm
Let the $ K'$ frame move with
$\vec{\beta}_{\perp 0} = \vec{P}_{\perp 0}/M_{\perp }$ velocity and
with Lorentz-factor $\gamma_{\perp 0}=E_{\perp 0}/M=M_{\perp }/M$ ,
where $E_{\perp 0}$ is energy of string in $K_{L}$ frame.
Thereby we obtain the $K_{rest}$ rest frame of the string.

\vskip 0.8 cm
{\em Transformation} from $K_{rest}$ to $K_{exchange}$ frame where the
$z_{exchange}$ axis is along one of two parton momentum vectors which are
before gluon exchange occurs :
\vskip 0.2 cm
The momentum vectors of partons are transformed from $K_{0}$ to $K_{rest}$
frame. The longitudinal $z_{rest}$ axis of $K_{rest}$ frame is parallel to the
longitudinal $z_{0}$ axis of $K_{0}$ c.m.s. of both collising particles.
We turn the $K_{rest}$ frame to obtain the $z_{exchange}$ axis along to the
momentum vector of one of two partons and the $y_{exchange}$ axis along to the
vector product $\vec{z}_{exchange} \times \vec{z}_{rest}$. Thereby we obtain
the $K_{exchange}$ frame\footnote{The parameters for transformation of momentum
projections from $K_{exchange}$ to $K_{rest}$ frame are given in Appendix 2.}
where we can determine the new parton momentum
vectors arising after gluon exchange, which leads to transition of two
partons on the mass shell (see Sect.1).

\vskip 0.8 cm
At the transition of both partons on the mass shell the modulus of momentum
of the parton in the $K_{exchange}$ frame changes. The energies of the partons
in the rest frame of string $K_{exchange}$ become equal to

\begin{equation}
\varepsilon_{1}^{exchange}= \frac{M^{2} + \mu^{2}_{1} - \mu^{2}_{2}}{2 M},
\;\;\;\;\;\;\;\;\;
\varepsilon_{2}^{exchange}= \frac{M^{2} + \mu^{2}_{2} - \mu^{2}_{1}}{2 M},
\end{equation}
where $ \mu_{1}$ and $\mu_{2}$ are the masses of partons. The diquark masses:
$m_{du}=0.45 GeV$, $m_{us}=0.62 GeV$,
$m_{ss}=0.79 GeV$ \cite{AnderGustIngSjos} are used.

As it has been discussed above in Sect.1, the Gauss parameter should be
connected with space like virtuality of parton, for example it might be the law
similar to one given by (2). However, in our calculations we have taken an
approximation (5). It means that Gauss parameter should be connected with
flouver of parton which does not lie on the mass shell before gluon exchange.
Therefore the $p_{x \; 1}^{exchange}$ and $p_{y \; 1}^{exchange}$ components of
the transverse momentum of one of two partons in the rest frame of string
$K_{exchange}$ are generated according to Gauss law

\begin{equation}
r_{1}(r_{2}) = \sigma^{-1} (2\pi)^{-0.5}
\int_{-\infty}^{p_{x}^{exchange}(p_{y}^{exchange})}
 exp(-x^{2}/2\sigma^{2})
\;\;\; dx  \;\;  ,
\end{equation}
where $r_{i}$ is uniformly distributed random number in the range (0,1)
and $\sigma$ is parameter, which is $\sigma_{u(d)}$, $\sigma_{s}$ or
$\sigma_{diquark}$ from (5).

Now one can determine new longitudinal projections of the parton momentum
vectors in the rest frame of string $K_{exchange}$ and transform
the momentum vectors of both partons from $K_{exchange}$ to $K_{rest}$
(see Appendix 3).
\vskip 1.0 cm

\section{The decay of string}

We follow the models \cite{AnderGustIngSjos}-\cite{Amelin} and
suppose that the cascade breaking of string is described by the iteration
of decay of string into hadron and string (Fig.3), i.e. by the iteration of
conversion of (di)quark into hadron and (di)quark (see Fig.5) via vacuum
pair production.

As it has been discussed in Sect.1, we should
generate the transverse projections $p_{x}$ and $p_{y}$ of vacuum quark $q'$
in the system where the transverse momentum of parent quark\footnote{It
might be either of the two partons taken with equal probability.}  $q$
(Fig.1b) is zero. So, the kinematical characteristics of
both partons should be transformed from $K_{rest}$ to this "decay" frame.
The possible way to do that is :

\vskip 0.8 cm
{\em Transformation} from $K_{rest}$ to $K''$ frame where
$x''$ axis is along the transverse momentum vector
of quark $q$ in the $K_{rest}$ system :
\vskip 0.2 cm

In the $K_{rest}$ frame, the quark $q$ has transverse momentum
$\vec{p}_{\perp \; rest}$ with projections $p_{x}^{rest}$ and $p_{y}^{rest}$.
We turn the $K_{rest}$ frame (in the transverse $xy$ plane) to obtain
the $x_{rest}$ axis along the vector $\vec{p}_{\perp \; rest}$.
The parameters for this transformation are
$ \cos\varphi = p_{x}^{rest}/p_{\perp \; rest} $,
$ \sin\varphi = p_{y}^{rest}/p_{\perp \; rest} $.

\vskip 0.8 cm
{\em Transformation} from $K''$ to $K_{decay}$ frame where the transverse
momentum of quark $q$ is equal zero :
\vskip 0.2 cm

Let the $ K''$ frame move with
$\vec{\beta}^{''}=\vec{p}_{\perp \; rest}/\varepsilon^{''}$ velocity where
$\varepsilon^{''}$ is the energy of quark $q$ in $K''$  ($K_{rest}$) frame.
Thereby we obtain the $K_{decay}$ frame, where the conversion
$q \rightarrow h + q'$ is generated and where the transverse
momentum of quark $q$ is equal zero, and so we are sure that
the functions, describing this conversion, can not depend on the transverse
momentum of quark $q$.

Three schemes of conversions of the (di)quark into hadron and (di)quark
are taking into consideration in our modeling (see Fig.5). The $a$ and
$b$ types of conversions is generated according to the relation
$a:b=0.85:0.15$.

The flavours of quarks, produced from vacuum at the decay of string, are
generated according to the relation (6) \cite{AnderGustIngSjos}.

In Monte Carlo simulation of decay (conversion) of a parton into another
parton and hadron (Fig.1b)
we have to take into account that fixed quark content of hadron might
correspond to a few types of particles. Therefore there are some decay
modes for the given quark content of hadron. We attribute a
weight to aech decay mode. This weight is equal to the product of two
factors. The spin factor is equal to $(2J+1)$, where $J$ is spin of the
hadron. The $SU_{3}$-factor is taken into consideration, if there are several
hadrons with the same quark content, spin and parity. For example,
$SU_{3}$-factor of $\eta$-meson which is formed
from $\overline{u}u$-pair is equal
to $1/6$ , and the same $SU_{3}$-factor of $w$-meson is equal to $1/2$.

If the hadron $h$ is resonanse, its mass is generated, after the generation
of the decay mode, according to the Breit-Wigner distribution.

In the $K_{decay}$ frame the $P_{x \; had}^{decay}$ and $P_{y \; had}^{decay}$
components of the transverse momentum $p_{\perp \; had}^{decay}$
of daughter hadron $h$ are generated according to the Gauss law (see Sect.1)
\begin{equation}
r_{1}(r_{2}) = \sigma^{-1} (2\pi)^{-0.5}
\int_{-\infty}^{P_{x \; had}^{decay}(P_{y \; had}^{decay})}
exp(-x^{2}/2\sigma^{2})   \;\;\; dx  \;\;\;\;\;\;  ,
\end{equation}
where $r_{i}$ is uniformly distributed random number in the range (0,1)
and $\sigma$ is parameter taken from (1).

At the convertion of parton $q$ (Figs.1b,3a) of the string $S$ (Fig.3a) into
daughter hadron $h$  in the $K_{decay}$ system  we generate the
convenient Lorentz-invariant variable
\begin{equation}
z = \frac{          E_{had}^{decay} + P_{z \; had}^{decay}}
         {\varepsilon_{parton}^{decay} + p_{z \; parton}^{decay}} \;\; .
\end{equation}
In (13) $ E_{had}^{decay}$ and $P_{z \; had}^{decay}$ are the energy and the
momentum projection (on $z_{decay}$ axis) of daughter hadron,
$\varepsilon_{parton}^{decay}$ and $p_{z \; parton}^{decay}$ are the same
values of the parton.

The distribution density of $z$ variable is parameterized by the form
\begin{equation}
F(z) \;\;   \propto \;\;  z \;\;\;\; , \;\;\;\; 0<z<1  \;\;\;  .
\end{equation}

After determination of the momentum and energy of $q'$ we know all kinematical
characteristics of both partons ($\bar{q}$ and $q'$) at the ends of daughter
string $S'$ (Fig.3a), and so we can find its energy $E'$ and momentum vector
$\vec{P'}$ and control the inequality
\begin{equation}
 E' \; > \; P' \;\; .
\end{equation}
If the inequality (15) is fulfilled for daughter string $S'$, we determine
its mass $M' = \sqrt{E'^{2} - P'^{2}}$ and control the inequality
\begin{equation}
 M' \; > \; \mu_{1} \; + \; \mu_{2} \;\;\; ,
\end{equation}
where $\mu_{1}$ and $\mu_{2}$ are fixed (di)quark masses
(taken from \cite{AnderGustIngSjos})  of the partons ($q'$ and $\bar{q}$
in Fig.3a), tensing the daughter string ($S'$ in Fig.3a).

If the inequality (16) is fulfilled, we control the possibility to be decay
of daughter string into two hadrons. So, the decay scheme of daughter string
$S'$ (Fig.6) into two hadrons and the flavours of quarks, produced in this decay
from vacuum are generated\footnote{The decay scheme (Fig.6) is generated
according to the relation $a \; : \; b \; = \; 0.85 \; : \; 0.15 \;\;$.}.
Then we control the inequality
\begin{equation}
 M' \; > \; M^{min}_{1} \; + \; M^{min}_{2} \;\;\; ,
\end{equation}
where $M^{min}_{1}$ and $M^{min}_{2}$ are minimum masses of the 1-st and 2-nd
hadrons with the fixed quark composition, which can be produced by decay of the
daughter string $S'$.

If the inequalities (15)-(17) are fulfilled, the energies and momentum vectors
of daughter hadron and partons of daughter string are transformed from the
$K_{decay}$ to $K_{0}$ frame.

If one of three inequalities (15)-(17) is not fulfilled, instead of the
decay of mother string $S$ into daughter string and hadron (Fig.3a), the decay
of mother string into two hadrons (Fig.6) is generated in the $K_{rest}$ frame.
In that case there are some  decay modes for the given quark content of these
hadrons as it was already  discussed above.

 The $M_{1}$ and $M_{2}$ masses of resonances are generated after
the generation of the decay mode. For example, if string decays
into two resonances and the mass of first resonance must be
generated at first, the $M_{1}$ and $M_{2}$ masses can be generated
according to the Breit-Wigner distributions in intervals
\begin{eqnarray}
  M^{min}_{1} \;\;  <  \;\; M_{1} \;\; < \;\; M - M^{min}_{2} \;\; ,   \nonumber
\end{eqnarray}
\begin{eqnarray}
  M^{min}_{2} \;\;  <  \;\; M_{2} \;\; <  \;\; M - M_{1} \;\; ,        \nonumber
\end{eqnarray}
where $M^{min}_{i}$ is maximum sum of masses of particles produced
by the decay of $i$-th resonance. In the $K_{rest}$ frame the square of
transverse momentum of resonance
$(P^{rest}_{\perp})^{2}=(P^{rest}_{x1})^{2} \; + \; (P^{rest}_{y1})^{2}$
is generated in the interval $(0,P^{2})$, and the azimuthal
angle is generated uniformly in the interval $(0,2\pi)$. The momentum of
resonance is supposed to have the sharp angle with momentum vector of
its parent parton. The decays of unstable hadrons into stable and
quasistable particles are generated,  for example, according to algorithm given
in the Appendix 2  from Ref. \cite{ChudLu} .
The momenta of stable and quasistable particles are transformed from
the $K_{rest}$ to $K_{0}$ frame.

The decay of the daughter string is generated also as the decay of the
mother string. The cascade is generated up to the decay of all strings
into hadrons.

\vskip 0.5 cm

\section{Conclusion}

The string decay is based on the vacuum pair production which could be
understood as a result of the confinement interaction. A quantitative
description for quark confinement was given in the models
\cite{Artru}-\cite{Werner} of muliparticle production where
the quark-antiquark colourless system is confined
by constant magnitude force which determines the breakpoint of string.
Here we try to constract a possible scenario for quantitative description of
the quark confinement interaction which is reason for vacuum pair production in
the widely  explored process of conversion of quark $q$ into hadron $h$ and
quark $q'$ (Fig.1b).   In other words, instead of the process described by
the diagram\footnote{It is an independent conversion of both partons $q$
and $\bar{q}$ at the ends of string into two jets of hadrons $h$.}
in Fig.1d  we use the diagram in Fig.3 where we add an
interaction (marked by the circle) to the process of coversion
$q \rightarrow h + q'$ which is seemed as a result of the confinement
interaction. To describe the confinement interaction we try to use an
essential idea of the parton model which regards a high energy
claster (hadron,string) as a collection of quasifree partons , and
we try to follow some concepts of the string model formulated in
\cite{AnderNils} where the gluons move like localised particles carrying energy
momentum between quark and antiquark at the
endpoints of string, and so we take into account they go on the mass shell
before the break of daughter string occurs (Fig.3). In the frame of that
scenario there is not any problem connected with virtuality\footnote{The quark
$q'$ is produced from vacuum and so even if a quark $q$ (Fig.1b) is on
the mass shell, the quark $q'$ has wide distributed spacelike virtuality
(see Fig.2) in its conversion $q' \rightarrow h + q''$ in the models based on
the diagram shown in Fig.1d.}
and simultaneously we try to take into account the interaction
between the partons at the ends of each string (colourless system).
At the gluon exchange the gluons carry also transverse momentum which could be
connected with spacelike virtuality of quark (Fig.2) at the end of string.
To describe the transverse $x$- and $y$-projections of parton momentum
at string rest frame (see Sect.3) after gluon exchange we choose\footnote{See
Sect.1} the Gauss law and an approximation (5).

Some results of Monte Carlo calculations for antiproton-proton interactions are
given in Fig.7, where the transverse momentum distribution for
the secondary Monte Carlo
particles depends on the energy of interaction. It is because of kinematical
correlations between the ends of string (confinement interaction). Namely,
in frame of presented scenario, there are two Gauss laws which describe
the distributions for the transverse momentum projections in two different
processes. The first Gauss law is applyed to the processes given in Figs.5,6
(the vacuum  pair production).  The second one is applyed to the
gluon exchange (circle in Fig.3). At the small energy of interaction of primary
proton and antiproton the most part of number of all primary strings (Fig.4)
have the decays directly into two hadrons (Fig.6). If the c.m. energy
 $\sqrt{s}$ grows, the average number of gluon exchanges (confinement
interactions) grows with the number of cascade  decays (Fig.3) of each primary
string (Fig.4). According to the approximation (5), that means, more
big $p_{T}$ of particles occur.

\vskip 0.8cm

\begin{center}

ACKNOWLEDGMENTS

\end{center}
The author would like to express his sincere thanks to
V.M.Chudakov for many discussions.

\vskip 1.5cm
{\large  \bf Appendix 1}
\vskip 0.4cm

The inelastic nondiffractive cross section corresponding to formation of
$l$ pairs of primary quark-gluon strings at the energy
$\sqrt{s}$ in c.m.s. has been taken from \cite{Kai-Ter-Mar} :
\begin{eqnarray}
\sigma_{l} = \frac{\sigma_{P}}{l z} ( 1 - \exp(-z)  \sum_{k=0}^{l-1} \frac{z^{k}}{k!} ) \;\;\;\;\;\;,  \nonumber
\end{eqnarray}
where
$ \sigma_{P} = 8 \pi \gamma \exp(\varepsilon \Delta) $ ,
$ \gamma = 3.64 GeV^{-2}     $  ,
$ \varepsilon = ln (s/s_{0}) $  ,
$ s_{0} = 1 GeV^{2}          $  ,
$ \Delta = 0.068             $  ,
$ z = 2 c \gamma \exp(\varepsilon \Delta) / (R^{2} + \alpha'_{P} \varepsilon ) $ ,
$ c = 1.5                    $  ,
$ R^{2} = 3.56 GeV^{-2}      $  ,
$ \alpha'_{P} = 0.25 GeV^{-2} $ .

\vskip 0.4cm
In calculation of probabilities $ W_{l} = \sigma_{l} / \sigma $ the inelastic
nondiffractive cross section $\sigma$ has been determined by formula
\begin{eqnarray}
\sigma = \sigma_{P} \sum_{\nu = 1}^{\infty} \frac{(-z)^{\nu-1}}{\nu\cdot\nu!} \;\;\;\; , \nonumber
\end{eqnarray}
and at any energy $\sqrt{s}$ in c.m.s. it is taken into account up to $l$=15
pairs of primary strings.

\vskip 1.5cm
{\large  \bf Appendix 2}
\vskip 0.4cm

The parameters to transformation of momentum projections from $K_{exchange}$ to
$K_{rest}$ frame are

\begin{eqnarray}
cos\theta  = \frac{ - p_{z}^{rest}}{ \sqrt{(p_{x}^{rest})^{2} + (p_{y}^{rest})^{2} +  + (p_{z}^{rest})^{2}}} , \nonumber
\end{eqnarray}

\begin{eqnarray}
cos\varphi = \frac{ - p_{x}^{rest} }{ \sqrt{(p_{x}^{rest})^{2} + (p_{y}^{rest})^{2}}} \; ,   \nonumber
\end{eqnarray}

\begin{eqnarray}
sin\varphi = \frac{ - p_{y}^{rest} }{ \sqrt{(p_{x}^{rest})^{2} + (p_{y}^{rest})^{2}}} \; ,   \nonumber
\end{eqnarray}
where $p_{x}^{rest}$, $p_{y}^{rest}$, $p_{z}^{rest}$  are the momentum
projections (in the $K^{rest}$ frame) of one of two partons  before
gluon exchange.

\vskip 1.5cm
{\large  \bf Appendix 3}
\vskip 0.4cm

New parton momentum projections, arising after gluon exchange, are transformed
from the $K_{exchange}$ to $K_{rest}$ frame according to formulas
\begin{eqnarray}
(p_{x}^{rest})_{new} = - p_{x}^{exchange}  cos\theta  cos\varphi
               - p_{y}^{exchange}  sin\varphi
               - p_{z}^{exchange}  sin\theta  cos\varphi ,              \nonumber
\end{eqnarray}
\begin{eqnarray}
(p_{z}^{rest})_{new}   =   p_{x}^{exchange} \; sin\theta
                 -p_{z}^{exchange} \; cos\theta   \;\;\;\;\; ,          \nonumber
\end{eqnarray}
\begin{eqnarray}
(p_{y}^{rest})_{new}   = - p_{x}^{exchange}  cos\theta sin\varphi
                 + p_{y}^{exchange}  cos\varphi
                 - p_{z}^{exchange}  sin\theta  sin\varphi  ,           \nonumber
\end{eqnarray}
where $p_{x}^{exchange}$, $p_{y}^{exchange}$, $p_{z}^{exchange}$ are the
momentum projections of one of two partons arising after gluon exchange
in the $K_{exchange}$ frame , and $\theta$, $\varphi$ angles are determined
in Appendix 2.

\newpage
\begin{center}
{\large  \bf Figure Captions}
\end{center}
\vskip 0.7 cm

{\large \bf Fig.1. } {\large \bf a} The diagram of conversion of the fast quark
      $q$ into  hadrons $h$ in the models \cite{AnderGustIngSjos}-\cite{Amelin}.
      {\large \bf b}. An elementary Monte Carlo process of the
      models \cite{AnderGustIngSjos}-\cite{Amelin}:  $q \rightarrow h + q'$ .
      {\large \bf c}. An elementary Monte Carlo process $q'\rightarrow h + q''$.
      {\large \bf d}. The diagram of independent conversion of the
      quark and antiquark at the ends of string into two jets of
      hadrons in the models  \cite{AnderGustIngSjos}-\cite{Amelin}.
      The double line is the diquark.

\vskip 0.8 cm
{\large \bf Fig.2 }
{\large \bf a},{\large \bf b}. The distributions for the mass squared invariant
$m_{q'}^{* 2} = E^{2}_{q'}- \vec{P}^{2}_{q'}$ of
$q'_{u}$-quark ({\large \bf a}) , $q'_{s}$-quark ({\large \bf b}) produced
in the process $q \rightarrow h + q'$,
if quark $q$ (see Figs.1b,5a) lies on the mass shell in each its conversion
into daughter hadron $h$ and quark $q'$.
{\large \bf c}. The distribution for the mass squared invariant
$m_{qq'}^{* 2} = E^{2}_{qq'}- \vec{P}^{2}_{qq'}$ of diquark $qq'$
produced in the process presented in Fig.5b,
if quark $q$ lies on the mass shell in each its conversion
into daughter hadron $h$ and diquark $qq'$.

\vskip 0.8 cm
{\large \bf Fig.3 } {\large \bf a}  The diagram of the gluon exchange
             (marked by circle) between  the quark $q$ and antiquark
              $\overline{q}$ at the ends of secondary string $S$ which
             then has decay into daughter hadron $h$ and string $S'$ via
             conversion $q \rightarrow h + q'$.   {\large \bf b}
             The diagram of the process $S' \rightarrow  S'' + h$ which is like
             one given in {\large \bf a} .

\vskip 0.8 cm
{\large \bf Fig.4 } The formation of four ($l$=2) primary strings $S_{1}$,
                    $S_{2}$, $S_{3}$, $S_{4}$ by partons of projectile and
                    target in inelastic $\overline{p}p$-interaction.
                    The double line is the diquark.

\vskip 0.8 cm
{\large \bf Fig.5 } Three schemes of conversion of the
                    quark into meson and quark ({\large \bf a}),
                    quark into baryon  and diquark ({\large \bf b}),
                    diquark into baryon and quark ({\large \bf c}),
                    via vacuum pair production.
                    The double line is the diquark.

\vskip 0.8 cm
{\large\bf Fig.6} Three schemes of decay of string into two daughter hadrons
                  $h_{1}$ and $h_{2}$ via vacuum pair production.
                  The double line is the diquark.

\vskip 0.8 cm
{\large \bf Fig.7 } Transverce momentum distribution for charged particles
                    produced in $\overline{p}p$-interactions at
                    $\sqrt{s}$=546 GeV. Experimental data are from
                    \cite{AlnerGJ}.

\end{document}